
\magnification=\magstephalf
\hsize=14.0 true cm
\vsize=19 true cm
\hoffset=1.0 true cm
\voffset=2.0 true cm

\abovedisplayskip=12pt plus 3pt minus 3pt
\belowdisplayskip=12pt plus 3pt minus 3pt
\parindent=1em


\font\sixrm=cmr6
\font\eightrm=cmr8
\font\ninerm=cmr9

\font\sixi=cmmi6
\font\eighti=cmmi8
\font\ninei=cmmi9

\font\sixsy=cmsy6
\font\eightsy=cmsy8
\font\ninesy=cmsy9

\font\sixbf=cmbx6
\font\eightbf=cmbx8
\font\ninebf=cmbx9

\font\eightit=cmti8
\font\nineit=cmti9

\font\eightsl=cmsl8
\font\ninesl=cmsl9

\font\sixss=cmss8 at 8 true pt
\font\sevenss=cmss9 at 9 true pt
\font\eightss=cmss8
\font\niness=cmss9
\font\tenss=cmss10

\font\bigrm=cmr10 at 12 true pt
\font\bigbf=cmbx10 at 12 true pt

\catcode`@=11
\newfam\ssfam

\def\tenpoint{\def\rm{\fam0\tenrm}%
    \textfont0=\tenrm \scriptfont0=\sevenrm \scriptscriptfont0=\fiverm
    \textfont1=\teni  \scriptfont1=\seveni  \scriptscriptfont1=\fivei
    \textfont2=\tensy \scriptfont2=\sevensy \scriptscriptfont2=\fivesy
    \textfont3=\tenex \scriptfont3=\tenex   \scriptscriptfont3=\tenex
    \textfont\itfam=\tenit                  \def\it{\fam\itfam\tenit}%
    \textfont\slfam=\tensl                  \def\sl{\fam\slfam\tensl}%
    \textfont\bffam=\tenbf \scriptfont\bffam=\sevenbf
    \scriptscriptfont\bffam=\fivebf
                                            \def\bf{\fam\bffam\tenbf}%
    \textfont\ssfam=\tenss \scriptfont\ssfam=\sevenss
    \scriptscriptfont\ssfam=\sevenss
                                            \def\ss{\fam\ssfam\tenss}%
    \normalbaselineskip=13pt
    \setbox\strutbox=\hbox{\vrule height8.5pt depth3.5pt width0pt}%
    \let\big=\tenbig
    \normalbaselines\rm}

\def\ninepoint{\def\rm{\fam0\ninerm}%
    \textfont0=\ninerm      \scriptfont0=\sixrm
                            \scriptscriptfont0=\fiverm
    \textfont1=\ninei       \scriptfont1=\sixi
                            \scriptscriptfont1=\fivei
    \textfont2=\ninesy      \scriptfont2=\sixsy
                            \scriptscriptfont2=\fivesy
    \textfont3=\tenex       \scriptfont3=\tenex
                            \scriptscriptfont3=\tenex
    \textfont\itfam=\nineit \def\it{\fam\itfam\nineit}%
    \textfont\slfam=\ninesl \def\sl{\fam\slfam\ninesl}%
    \textfont\bffam=\ninebf \scriptfont\bffam=\sixbf
                            \scriptscriptfont\bffam=\fivebf
                            \def\bf{\fam\bffam\ninebf}%
    \textfont\ssfam=\niness \scriptfont\ssfam=\sixss
                            \scriptscriptfont\ssfam=\sixss
                            \def\ss{\fam\ssfam\niness}%
    \normalbaselineskip=12pt
    \setbox\strutbox=\hbox{\vrule height8.0pt depth3.0pt width0pt}%
    \let\big=\ninebig
    \normalbaselines\rm}

\def\eightpoint{\def\rm{\fam0\eightrm}%
    \textfont0=\eightrm      \scriptfont0=\sixrm
                             \scriptscriptfont0=\fiverm
    \textfont1=\eighti       \scriptfont1=\sixi
                             \scriptscriptfont1=\fivei
    \textfont2=\eightsy      \scriptfont2=\sixsy
                             \scriptscriptfont2=\fivesy
    \textfont3=\tenex        \scriptfont3=\tenex
                             \scriptscriptfont3=\tenex
    \textfont\itfam=\eightit \def\it{\fam\itfam\eightit}%
    \textfont\slfam=\eightsl \def\sl{\fam\slfam\eightsl}%
    \textfont\bffam=\eightbf \scriptfont\bffam=\sixbf
                             \scriptscriptfont\bffam=\fivebf
                             \def\bf{\fam\bffam\eightbf}%
    \textfont\ssfam=\eightss \scriptfont\ssfam=\sixss
                             \scriptscriptfont\ssfam=\sixss
                             \def\ss{\fam\ssfam\eightss}%
    \normalbaselineskip=10pt
    \setbox\strutbox=\hbox{\vrule height7.0pt depth2.0pt width0pt}%
    \let\big=\eightbig
    \normalbaselines\rm}

\def\tenbig#1{{\hbox{$\left#1\vbox to8.5pt{}\right.\n@space$}}}
\def\ninebig#1{{\hbox{$\textfont0=\tenrm\textfont2=\tensy
                       \left#1\vbox to7.25pt{}\right.\n@space$}}}
\def\eightbig#1{{\hbox{$\textfont0=\ninerm\textfont2=\ninesy
                       \left#1\vbox to6.5pt{}\right.\n@space$}}}

\font\sectionfont=cmbx10
\font\subsectionfont=cmti10

\def\figurecaptionfont{\ninepoint}
\def\tablecaptionfont{\ninepoint}
\def\footnotefont{\eightpoint}


\newcount\equationno
\newcount\bibitemno
\newcount\figureno
\newcount\tableno

\equationno=0
\bibitemno=0
\figureno=0
\tableno=0
\advance\pageno by -1


\footline={\ifnum\pageno=0{\hfil}\else
{\hss\rm\the\pageno\hss}\fi}


\def\section #1 \par
{\vskip3ex
\global\def\equationlabel{#1}
\global\equationno=0
\immediate\write\terminal{Section #1.}}


\def\subsection #1 \par
{\vskip0pt plus .15\vsize\penalty-50 \vskip0pt plus-.15\vsize
\vskip2.5ex plus 0.1ex minus 0.1ex
\leftline{\subsectionfont #1}\par
\immediate\write\terminal{Subsection #1}
\vskip1.0ex plus 0.1ex minus 0.1ex
\noindent}


\def\appendix #1 \par
{\vskip3ex
\global\def\equationlabel{\hbox{\rm#1}}
\global\equationno=0
\immediate\write\terminal{Appendix #1}}


\def\enum{\global\advance\equationno by 1
(\equationlabel.\the\equationno)}


\def\ifundefined#1{\expandafter\ifx\csname#1\endcsname\relax}

\def\ref#1{\ifundefined{#1}?\immediate\write\terminal{unknown reference
on page \the\pageno}\else\csname#1\endcsname\fi}

\newwrite\terminal
\newwrite\bibitemlist

\def\bibitem#1#2\par{\global\advance\bibitemno by 1
\immediate\write\bibitemlist{\string\def
\expandafter\string\csname#1\endcsname
{\the\bibitemno}}
\item{[\the\bibitemno]}#2\par}

\def\beginbibliography{
\vskip0pt plus .20\vsize\penalty-150 \vskip0pt plus-.20\vsize
\vskip 1.6 true cm plus 0.2 true cm minus 0.2 true cm
\centerline{\sectionfont References}\par
\immediate\write\terminal{References}
\immediate\openout\bibitemlist=biblist
\frenchspacing
\vskip 0.7 true cm plus 0.1 true cm minus 0.1 true cm}

\def\endbibliography{
\immediate\closeout\bibitemlist
\nonfrenchspacing}


\def\figurecaption#1{\global\advance\figureno by 1
\narrower\figurecaptionfont
Fig.~\the\figureno. #1}

\def\tablecaption#1{\global\advance\tableno by 1
\vbox to 0.5 true cm { }
\centerline{\tablecaptionfont%
Table~\the\tableno. #1}
\vskip-0.4 true cm}

\tenpoint

\immediate\openin\bibitemlist=biblist
\ifeof\bibitemlist\immediate\closein\bibitemlist
\else\immediate\closein\bibitemlist
\input biblist \fi


\def\blackboardrrm{\mathchoice
{\rm I\kern-0.21 em{R}}{\rm I\kern-0.21 em{R}}
{\rm I\kern-0.19 em{R}}{\rm I\kern-0.19 em{R}}}

\def\blackboardzrm{\mathchoice
{\rm Z\kern-0.32 em{Z}}{\rm Z\kern-0.32 em{Z}}
{\rm Z\kern-0.28 em{Z}}{\rm Z\kern-0.28 em{Z}}}

\def\blackboardh{\mathchoice
{\ss I\kern-0.14 em{H}}{\ss I\kern-0.14 em{H}}
{\ss I\kern-0.11 em{H}}{\ss I\kern-0.11 em{H}}}

\def\blackboardp{\mathchoice
{\ss I\kern-0.14 em{P}}{\ss I\kern-0.14 em{P}}
{\ss I\kern-0.11 em{P}}{\ss I\kern-0.11 em{P}}}

\def\blackboardt{\mathchoice
{\ss T\kern-0.52 em{T}}{\ss T\kern-0.52 em{T}}
{\ss T\kern-0.40 em{T}}{\ss T\kern-0.40 em{T}}}


\def\thisyear{\number\year}

\def\thismonth{\ifcase\month\or
January\or February\or March\or April\or May\or June\or
July\or August\or September\or October\or November\or December\fi}



\def\rmd{{\rm d}}

\def\rme{{\rm e}}



\def\proof{\noindent{\sl Proof:}\kern0.6em}

\def\frac#1#2{\hbox{$#1\over#2$}}
\def\dual{\mathstrut^*\kern-0.1em}

\def\ring{\mathaccent"7017}
\def\lvec#1{\setbox0=\hbox{$#1$}
    \setbox1=\hbox{$\scriptstyle\leftarrow$}
    #1\kern-\wd0\smash{
    \raise\ht0\hbox{$\raise1pt\hbox{$\scriptstyle\leftarrow$}$}}
    \kern-\wd1\kern\wd0}
\def\rvec#1{\setbox0=\hbox{$#1$}
    \setbox1=\hbox{$\scriptstyle\rightarrow$}
    #1\kern-\wd0\smash{
    \raise\ht0\hbox{$\raise1pt\hbox{$\scriptstyle\rightarrow$}$}}
    \kern-\wd1\kern\wd0}


\def\nabstar#1{\nabla\kern-0.5pt\smash{\raise 4.5pt\hbox{$\ast$}}
               \kern-4.5pt_{#1}}
\def\drv#1{{\partial_{#1}}}
\def\drvstar#1{\partial\kern-0.5pt\smash{\raise 4.5pt\hbox{$\ast$}}
               \kern-5.0pt_{#1}}




\def\psibar{\overline{\psi}}

\def\chibar{\overline{\chi}}


\def\dirac#1{\gamma_{#1}}
\def\diracstar#1#2{
    \setbox0=\hbox{$\gamma$}\setbox1=\hbox{$\gamma_{#1}$}
    \gamma_{#1}\kern-\wd1\kern\wd0
    \smash{\raise4.5pt\hbox{$\scriptstyle#2$}}}


\def\tr{\,\hbox{tr}\,}


\def\Sf{S_{\rm F}}
\def\Dtilde{\widetilde{D}}
\def\deltatilde{\tilde{\delta}\kern1pt}

\rightline{DESY 98-014}

\vskip 3.0 true cm minus 0.3 true cm
\centerline
{\bigbf Exact chiral symmetry on the lattice and}
\vskip1ex
\centerline
{\bigbf the Ginsparg-Wilson relation} 
\vskip 1.5 true cm
\centerline{\bigrm Martin L\"uscher}
\vskip1ex
\centerline{\it Deutsches Elektronen-Synchrotron DESY}
\centerline{\it Notkestrasse 85, D-22603 Hamburg, Germany}
\centerline{\it E-mail: luscher@mail.desy.de}
\vskip 3.0 true cm
\centerline{\bf Abstract}
\vskip 1.5ex
It is shown that the Ginsparg-Wilson relation implies
an exact symmetry of the fermion action, which may be regarded as 
a lattice form of an infinitesimal chiral rotation.
Using this result it is straightforward to construct lattice Yukawa
models with unbroken flavour and chiral symmetries
and no doubling of the fermion spectrum.
A contradiction with the Nielsen-Ninomiya theorem is avoided,
because the chiral symmetry is realized in a different way 
than has been assumed when proving the theorem.
\vfill
\centerline{\thismonth\space\thisyear}
\eject

\section 1 

{\bf 1.}~A well-known problem with fermions on the lattice is 
that one usually ends up with breaking chiral symmetry or having
more particles in the continuum limit than intended.
The celebrated Nielsen-Ninomiya theorem [\ref{NN}] states
that this is in fact unavoidable if a few plausible
assumptions are made. The construction of 
chiral field theories on the lattice
thus appears to be difficult and 
maybe even impossible in some cases.

Recently some intriguing results have been published by 
Neuberger [\ref{NeubergerI},\ref{NeubergerII}] and by 
Hasenfratz, Laliena and Niedermayer [\ref{HasenfratzEtAl}],
which suggest that chiral symmetry may be preserved 
in lattice QCD, at least to some 
extent, if the lattice Dirac operator is of a particular form.
The proposed expressions for the Dirac operator have been 
derived in completely different ways and tend to be  
very complicated, but all of them
satisfy a simple identity, originally due to
Ginsparg and Wilson [\ref{GinspargWilson}],
which protects the quark masses from additive renormalizations
[\ref{NeubergerI},\ref{HasenfratzNeu}]
and which plays a key r\^ole in the proof of the
lattice index theorem of ref.~[\ref{HasenfratzEtAl}].

In this letter it will be shown 
that the lattice fermion action in fact has
an exact symmetry if the Ginsparg-Wilson identity holds.
The usefulness of this observation is illustrated by
constructing a class of chiral Yukawa models on the lattice
with unbroken chiral and flavour symmetries (and no doublers).
Since the chiral transformations in these theories 
are not of the naively expected form,
a contradiction with the Nielsen-Ninomiya theorem is avoided 
without having to compromise in any other ways.
In particular, the flavour-singlet chiral symmetry 
has the expected anomaly if gauge interactions are included.

\section 2

{\bf 2.}~A particularly simple form of the Nielsen-Ninomiya theorem holds 
for free Dirac fermions on a euclidean lattice and 
it is helpful for the discussion that follows
to briefly recall this.
So let us consider the free field action 
$$
  \Sf=a^4\sum_x\;\psibar D\psi,
  \eqno\enum
$$
where $a$ denotes the lattice spacing and $D$ the lattice Dirac operator.
As usual we assume $D$ to be invariant under translations so that
$$
  D\rme^{ipx}u=\Dtilde(p)\rme^{ipx}u
  \eqno\enum
$$
for all constant Dirac spinors $u$ and some 
complex $4\times4$ matrix $\Dtilde(p)$.
The theorem now states that the following properties cannot
hold simultaneously (for an elegant proof see ref.~[\ref{Friedan}]).

\vskip1.5ex
\hskip0.25 true cm
\vbox{\hsize=13.0 true cm
\item{(a)}{$\Dtilde(p)$ is an analytic periodic function of the momenta 
$p_{\mu}$ with period $2\pi/a$.}}

\vskip1.0ex
\hskip0.25 true cm
\vbox{\hsize=13.0 true cm
\item{(b)}{For momenta far below the cutoff $\pi/a$, 
we have $\Dtilde(p)=i\dirac{\mu}p_{\mu}$
up to terms of order $ap^2$.}}

\vskip1.0ex
\hskip0.25 true cm
\vbox{\hsize=13.0 true cm
\item{(c)}{$\Dtilde(p)$ is invertible at all non-zero momenta 
(mod $2\pi/a$).}} 

\vskip1.0ex
\hskip0.25 true cm
\vbox{\hsize=13.0 true cm
\item{(d)}{$D$ anti-commutes with $\dirac{5}$.}}

\vskip1.0ex
\noindent
Property (a) is necessary if we want $D$ to be an essentially local
operator, (b) and (c) ensure that the correct continuum limit
is obtained and (d) guarantees that the fermion action 
is invariant under continuous chiral transformations
\footnote{$\dagger$}{\footnotefont
The Dirac matrices are taken to be hermitean and
$\dirac{5}=\dirac{0}\dirac{1}\dirac{2}\dirac{3}$}.

\section 3

{\bf 3.}~To escape the theorem,
Ginsparg and Wilson [\ref{GinspargWilson}] suggested many years ago
to replace property (d) through the relation
$$
  \dirac{5}D+D\dirac{5}=aD\dirac{5}D.
  \eqno\enum
$$
A simple consequence of this equation is
that the fermion propagator anti-commutes with $\dirac{5}$ 
at non-zero distances and chiral symmetry is thus partly preserved. 

Examples of free lattice Dirac operators satisfying 
the Ginsparg-Wilson relation can be found rather easily. 
A particularly simple solution is given by [\ref{NeubergerI}]
$$
  D={1\over a}\bigl\{1-A(A^{\dagger}A)^{-1/2}\bigr\},
  \qquad
  A=1-aD_{\rm w},
  \eqno\enum
$$
where $D_{\rm w}$ denotes the standard Wilson-Dirac operator, 
$$
  D_{\rm w}=\frac{1}{2}\left\{\dirac{\mu}(\drvstar{\mu}+\drv{\mu})
  -a\drvstar{\mu}\drv{\mu}\right\},
  \eqno\enum
$$
and $\drv{\mu}$ and $\drvstar{\mu}$ are the nearest-neighbour 
forward and backward difference operators.
Because of the square root in eq.~(3.2),
one might assume that $D$ is a non-local operator,
but this is actually not the case.
Using the abbreviations $\ring{p}_{\mu}=(1/a)\sin(ap_{\mu})$ and 
$\hat{p}_{\mu}=(2/a)\sin(ap_{\mu}/2)$, we have
$$
  a\Dtilde(p)=
  1-\Bigl\{1-\frac{1}{2}a^2\hat{p}^2-ia\dirac{\mu}\ring{p}_{\mu}\Bigr\}
  \Bigl\{1+\frac{1}{2}a^4
  \sum_{\mu<\nu}\hat{p}^2_{\mu}\hat{p}^2_{\nu}\Bigr\}^{-1/2},
  \eqno\enum
$$
and it is immediately clear from this formula that 
the conditions (a), (b) and (c) 
listed above are fulfilled.
In particular, from the analyticity of $\Dtilde(p)$ one infers
that its Fourier transform falls off exponentially at large distances
with a rate proportional to $1/a$. 
For free fermions eq.~(3.2) thus provides 
a completely satisfactory solution of the Ginsparg-Wilson relation.

\section 4

{\bf 4.}~We now show that eq.~(3.1) implies a
continuous symmetry of the fermion action, 
which may be interpreted as a lattice form of chiral symmetry.
No particular assumptions need to be made here, i.e.~the 
discussion applies to any 
Dirac operator satisfying the Ginsparg-Wilson
relation, including the gauge covariant operators 
of refs.~[\ref{NeubergerI}--\ref{HasenfratzEtAl}].

The infinitesimal variation of the fields associated with the
new symmetry is 
$$
  \delta\psi=\dirac{5}\left(1-\frac{1}{2}aD\right)\psi,
  \qquad
  \delta\psibar=\psibar\left(1-\frac{1}{2}aD\right)\dirac{5},
  \eqno\enum
$$
where $D$ is considered to be a matrix which
may be multiplied from the right 
with $\psi$ or from the left with $\psibar$.
Flavour non-singlet chiral transformations may be defined 
similarly by including a group generator in eq.~(4.1).
In both cases it is trivial to check that 
the fermion action eq.~(2.1) is invariant
if the Ginsparg-Wilson identity holds.

The flavour-singlet chiral symmetry 
is anomalous in the presence of gauge fields and it now seems that
we have got too much symmetry on the lattice.
The paradox is resolved by noting that
the fermion integration measure is in general not
invariant under the transformation (4.1). 

To work this out let us consider the theory in a finite space-time volume
with suitable boundary conditions so that the Ginsparg-Wilson identity
is preserved. We are then interested in the symmetry properties
of the (unnormalized) expectation values
$$
  \langle{\cal O}\rangle_{\rm F}=
  \int\prod_x\rmd\psi(x)\rmd\psibar(x)\,
  {\cal O}\,\rme^{-\Sf}
  \eqno\enum
$$
of arbitrary products $\cal O$ of the fermion fields.
By substituting 
$$
  \psi\to\psi+\epsilon\delta\psi,
  \qquad
  \psibar\to\psibar+\epsilon\delta\psibar,
$$
and expanding to first order in $\epsilon$ one obtains
$$
  \langle\delta{\cal O}\rangle_{\rm F}=
  -a\tr\{\dirac{5}D\}
  \langle{\cal O}\rangle_{\rm F},
  \eqno\enum
$$
where the trace is to be taken over the space of all fermion fields.
Evidently in the case of free fermions, with $D$ as given above, 
the trace vanishes and the symmetry is exact. The same is also
true if we consider flavour non-singlet chiral rotations, because
the group generator which has to be included in the transformation
law (and which then appears on the right-hand side of eq.~(4.3)) 
is traceless. 

The anomaly, $-a\tr\{\dirac{5}D\}$, has previously been calculated
in ref.~[\ref{HasenfratzEtAl}] and we now give a second derivation
which is applicable also in those cases where the Dirac operator 
does not have any particular hermiticity properties. 
Let $z$ be any complex number not contained in the spectrum of $D$.
A little algebra, using the Ginsparg-Wilson identity, yields
$$
  a(z-D)\dirac{5}(z-D)=z(2-az)\dirac{5}
  -(1-az)\left\{(z-D)\dirac{5}+\dirac{5}(z-D)\right\},
  \eqno\enum
$$
and after multiplying this equation from the right with $(z-D)^{-1}$
and taking the trace one ends up with
$$
  -a\tr\{\dirac{5}D\}=
  z(2-az)\tr\{\dirac{5}(z-D)^{-1}\}.
  \eqno\enum
$$
We now divide through the factor $z(2-az)$ and 
integrate over a small circle centred at the origin
that does not encircle any spectral value of $D$ other than $0$.
In particular,
$$
  P_0=\oint{\rmd z\over2\pi i}\,(z-D)^{-1}
  \eqno\enum
$$
projects on the subspace of zero modes of $D$ and
the result
$$
  -a\tr\{\dirac{5}D\}=
  2\tr{\dirac{5}P_0}=2N_{\rm f}\times\hbox{index}(D)
  \eqno\enum
$$
is thus obtained, where $N_{\rm f}$ denotes the number of fermion flavours.
Taken together eqs.~(4.3) and (4.7) show that the 
Ward identities associated with the global 
flavour-singlet chiral transformations on the lattice 
have the correct anomaly. 
In view of the exact index theorem 
of Hasenfratz et al.~[\ref{HasenfratzEtAl}]
and the earlier work of Ginsparg and Wilson~[\ref{GinspargWilson}]
this comes hardly as a surprise, but 
it is striking that the anomaly can be calculated with so little
effort.

\section 5

{\bf 5.}~It is now relatively easy to
couple fermions to scalar fields in such a way that 
the flavour and chiral symmetries are preserved on 
the lattice. Our starting point is the free fermion action 
$$
  \Sf=a^4\sum_x\,\bigl\{\psibar D\psi-(2/a)\chibar\chi\bigr\},
  \eqno\enum
$$
where $D$ is assumed to be a decent solution of 
the Ginsparg-Wilson relation such as the one discussed in section~3.
The auxiliary fields $\chi$ and $\chibar$ will later be used to construct 
chirally invariant interaction terms. 
For the time being we only note
that they do not propagate and the physical content of the theory
is hence unchanged.

As before one can show that the modified transformation 
$$
  \eqalign{
  &\delta\psi=\dirac{5}\left(1-\frac{1}{2}aD\right)\psi
  +\dirac{5}\chi, \quad\qquad 
  \delta\chi=\dirac{5}\frac{1}{2}aD\psi, \cr
  \noalign{\vskip1.5ex}
  &\delta\psibar=\psibar\left(1-\frac{1}{2}aD\right)\dirac{5}
  +\chibar\dirac{5}, \quad\qquad 
  \delta\chibar=\psibar\frac{1}{2}aD\dirac{5},\cr}
  \eqno\enum
$$
leaves the action and the fermion integration measure invariant
(gauge interactions are excluded in this section).
It follows from these equations that 
$$
  \delta(\psi+\chi)=\dirac{5}(\psi+\chi),
  \qquad
  \delta(\psibar+\chibar)=(\psibar+\chibar)\dirac{5},
  \eqno\enum
$$
and the propagator of the sum $\psi+\chi$ is hence chirally
invariant in the ordinary sense.
This is, incidentally, perfectly consistent with the 
Nielsen-Ninomiya theorem, because 
the Fourier transform of the propagator vanishes at some
momenta and its inverse is hence singular, thus 
violating property (a) (cf.~sect.~2).

Suppose now that $\phi$ is a complex
scalar field on the lattice with the usual 
self-interactions. A chirally invariant Yukawa interaction 
term is then given by
$$
  S_{\rm int}=
  a^4\sum_{x}\,g_0(\psibar+\chibar)\bigl\{
  \frac{1}{2}(1-\dirac{5})\phi+
  \frac{1}{2}(1+\dirac{5})\phi^{\ast}\bigr\}
  (\psi+\chi)
  \eqno\enum
$$
with $g_0$ being the bare coupling constant. 
More complicated interactions with flavour 
symmetries and various multiplets of fermions
can be constructed similarly.
A few remarks should be added at this point to make 
it clear that the lattice theories defined in this
way are completely sane. For simplicity 
attention is restricted to the
phase where chiral symmetry
is not spontaneously broken.

\vskip1ex
\noindent
(\romannumeral1)~In perturbation theory the one-particle irreducible diagrams
are chirally invariant in the ordinary sense, because 
the internal fermion lines represent the propagation
of $\psi+\chi$ and the vertices are manifestly invariant.
Non-symmetric counterterms are hence not needed to 
renormalize the theory.

\vskip1ex
\noindent
(\romannumeral2)~At sufficiently weak coupling the spectrum of fermions
is exactly as expected, i.e.~there are no doublers.
To see this first note that 
there are none when the interactions are switched off.
Now since the one-particle irreducible self-energy diagrams
anti-commute with $\dirac{5}$, they must be proportional to 
$\dirac{\mu}p_{\mu}$ at small momenta. In particular, 
the perturbative corrections to the fermion propagator are
of the form $D^{-1}B$ where $B$ is bounded
and it is hence impossible that new poles arise
at small couplings.

\vskip1ex
\noindent
(\romannumeral3)~The auxiliary fields $\chi$ and $\chibar$ 
couple only locally and do not carry any independent physical 
information. Essentially these fields play the r\^ole of 
Lagrange multipliers which may integrated out if so desired
although the expressions that one obtains are not particularly
illuminating
\footnote{$\dagger$}{\footnotefont
The bilinear part of the auxiliary field action can have
zero modes 
at large scalar fields. This does not lead to any singularities
(fermion integrals are always finite), but one may prefer
to avoid this complication by replacing the scalar field
in eq.~(5.4) through $\phi/(1+g_0^2a^2|\phi|^2)^{1/2}$.}.

\section 6

{\bf 6.}~The important qualitative message of this paper is 
that the Nielsen-Ninomiya theorem can be bypassed
if we do not insist that the chiral transformations assume
their canonical form on the lattice.
The construction of chirally invariant lattice theories
remains non-trivial, however, because the 
transformation laws depend on the interaction in general.
This is in fact required in gauge theories 
as otherwise one would end up with a non-anomalous 
flavour-singlet chiral symmetry.
An interesting observation in this connection is that 
the chiral rotations (5.2) become interaction dependent
when the auxiliary fields are eliminated.

At this point many interesting questions have
not even been touched and are left for future research.
In particular, the precise conditions under which 
chiral symmetries can exist on the lattice remain to be uncovered
and an attempt should be made to derive an identity, similar
to the Ginsparg-Wilson relation, which allows one
to construct exactly supersymmetric lattice theories.

\vskip1ex
Helpful correspondence with Karl Jansen, Peter Hasenfratz, Ferenc Niedermayer 
and Herbert Neuberger is gratefully acknowledged. 
I am also indebted to Istvan Montvay
for sharing his views on chiral Yukawa models.

\vfill\eject

\beginbibliography

\bibitem{NN}
N. B. Nielsen and M. Ninomiya,
Phys. Lett. B105 (1981) 219;
Nucl. Phys. B185 (1981) 20 [E: B195 (1982) 541];
{\it ibid}\/ B193 (1981) 173

\bibitem{NeubergerI}
H. Neuberger,
Exactly massless quarks on the lattice,
hep-lat/9707022

\bibitem{NeubergerII}
H. Neuberger,
More about exactly massless quarks on the lattice,
hep-lat/9801031

\bibitem{HasenfratzEtAl}
P. Hasenfratz, V. Laliena and F. Niedermayer,
The index theorem in QCD with a finite cut-off,
hep-lat/9801021

\bibitem{GinspargWilson}
P. H. Ginsparg and K. G. Wilson,
Phys. Rev. D25 (1982) 2649

\bibitem{HasenfratzNeu}
P. Hasenfratz,
Lattice QCD without tuning, mixing and current renormalization,
hep-lat/9802007

\bibitem{Friedan}
D. Friedan,
Commun. Math. Phys. 85 (1982) 481

\endbibliography
\vfill

\bye